# 1.06 μm Q-switched ytterbium-doped fiber laser using few-layer topological insulator $Bi_2Se_3$ as a saturable absorber

Zhengqian Luo[1,*], Yizhong Huang[1], Jian Weng[2,*], Huihui Cheng[1], Zhiqin Lin[2], Zhiping Cai[1] and Huiying Xu[1]

[1] *Department of Electronic Engineering, Xiamen University, Xiamen 361005, China*
[2] *Department of Biomaterials, College of Materials, Xiamen University, Xiamen 361005, China*
*Corresponding authors: zqluo@xmu.edu.cn, jweng@xmu.edu.cn*

**Abstract:** Passive Q-switching of an ytterbium-doped fiber (YDF) laser with few-layer topological insulator (TI) is, to the best of our knowledge, experimentally demonstrated for the first time. The few-layer TI: $Bi_2Se_3$ (2-4 layer thickness) is fabricated by the liquid-phase exfoliation method, and has a low saturable optical intensity of 53 $MW/cm^2$ measured by the Z-scan technique. The optical deposition technique is used to induce the few-layer TI in the solution onto a fiber ferrule for successfully constructing the fiber-integrated TI-based saturable absorber (SA). By inserting this SA into the YDF laser cavity, stable Q-switching operation at 1.06 μm is achieved. The Q-switched pulses have the shortest pulse duration of 1.95 μs, the maximum pulse energy of 17.9 nJ and a tunable pulse-repetition-rate from 8.3 to 29.1 kHz. Our results indicate that the TI as a SA is also available at 1 μm waveband, revealing its potential as another wavelength-independent SA (like graphene).

## 1. Introduction

Passively Q-switched fiber lasers with the advantages of compactness and flexibility have attracted much attention, and have been opening up some important applications in laser material processing, remote sensing, telecommunications and medicine. One of the most effective methods for passive Q-switching in fiber lasers is to use the saturable absorption of optical materials (i.e. saturable absorbers, SAs). In the past two decades, several kinds of SAs (e.g. semiconductor saturable absorber mirror SESAM [1], carbon nanotubes CNTs [2], graphene [3,4]) have been successively discovered, and successfully used to passively Q-switch fiber lasers. Even now, researchers are still making some efforts to seek for new SAs which are expected to have the ideal characteristics of wavelength-independent saturable absorption, low saturable optical intensity, high damage threshold and fiber compatibility. Most recently, similar to the graphene SA [5-8], the topological insulators (TIs) have been found to be another potential and ideal SAs for mode-locking or Q-switching [9-12].

TIs are states of quantum matter with the metallic states on the surface and have the narrow topological non-trivial energy gaps (e.g. TI: Bi2Se3, $\Delta E$~0.3 eV) [13], exhibiting the Dirac-like linear dispersion. When the narrow-bandgap TI is excited by strong light with the single-photon energy more than the TI bandgap $\Delta E$, the saturable absorption could happen due to the Pauli-blocking principle [9-11], like graphene. Therefore, TIs could also possess the ultra-broadband saturable absorption. For example, the saturable-absorption wavelength range of TI: Bi2Se3 is calculated to be from ultraviolet to mid-infrared <4.1 μm. Recently, Zhao et al. have measured that the saturable optical intensity of multilayer (>50 layers) TI:Bi2Se3 is 10.12 GW/cm2 at 800 nm [14] and 0.49 GW/cm$^2$ at 1550 nm [10]. By further inserting such TI-based SAs into the fiber laser cavities, their group has successfully mode-locked the Er-doped fiber lasers at 1.55 μm [9,10]. However, it should be noted that: 1) the broadband saturable-absorption characteristic of the TIs could not be fully exploited, 2) the TIs were not used to Q-switch fiber laser previously, and 3) if using few-layer TIs instead of multilayer TIs, its saturable-absorption performance could be more excellent and the corresponding mode-locking/Q-switching of fiber laser may be more favorable. In this letter, we report what we believe to be the first demonstration of a passively Q-switched ytterbium-doped fiber (YDF) laser with few-layer TI: Bi2Se3 as a SA. The Q-switched YDF laser operates in the 1.06 μm wavelength which is sufficiently away from the 1.55 μm waveband of TI-based Er-doped lasers reported previously [9-12], further demonstrating the broadband saturable-absorption nature of the TI.

## 2. Preparation of few-layer TI:Bi2Se3 and its optical absorption

The few-layer TI: Bi2Se3 used in our experiment was prepared by the liquid-phase exfoliation method as follows. Initially, we synthesized the Bi2Se3 crystal from bismuth oxide powder (Bi2O3) and selenium powder (Se) through the hydrothermal method. Then, the as-synthesized Bi2Se3 were added into the N-methyl-2-pyrrolidone (NMP) solution and sonicated for 24 h to produce the few-layer TI: Bi2Se3 suspension [the inset of Fig.1]. The thickness of the as-prepared TI nanosheets was characterized by atomic force microscopy (AFM), as shown in Fig. 1. The average thickness from the height profile diagram [Fig.1(b)] was measured to be ~2-4 nm. This indicates that the TI nanosheets are around 2-4 layers, because the single-layer thickness of Bi2Se3 is 0.96 nm [15]. It is interestingly noticed that the layer number (2-4 layer) of our TI: Bi2Se3 is much less than that (>50 layer) of TI: Bi2Se3 [10,14] previously reported by the hydrothermal intercalation/exfoliation method.

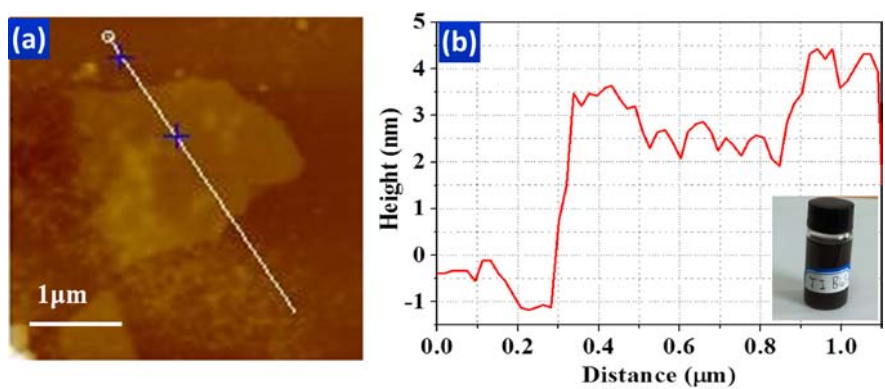

Fig.1. (a) The AFM image and (b) the height profile diagram of the few-layer TI: Bi2Se3. *Inset*: the as-prepared TI: Bi2Se3 suspension.

In order to conveniently measure the optical absorption properties of the few-layer TI: Bi2Se3, we sprayed the TI suspension onto a glass substrate by the spin-coating method, and then evaporating to dryness in an oven. As shown in Fig2.(a), we measured its transmission spectrum (i.e. linear-absorption spectrum) from 300 to 2000 nm using an optical spectrometer (Perkinelmer Lambda 7500). The linear-absorption curve is flat from ultraviolet to infrared region with 79.8% transmittance at 1.06 μm, indicating the potential of the few-layer TI as a broadband optical material. Using the Z-scan technique, we also characterized the saturable absorption of the TI sample. As described in Fig.2(b), the open-aperture Z-scan curve shows the real modulation depth ($\Delta T$) of 3.8% and the saturable optical intensity ($I_{sa}$) of 53 MW/cm$^2$ at 800 nm wavelength. Interestingly, the saturable intensity of our TI is about two-orders of magnitude less than the reported one (10.12 GW/cm$^2$) of the multilayer (>50 layer) TI [14], mainly benefiting from the few-layer (2-4 layers) structure. With such low saturable intensity, it can be predicted that the threshold of mode-locking or Q-switching using the few-layer TI can be significantly reduced.

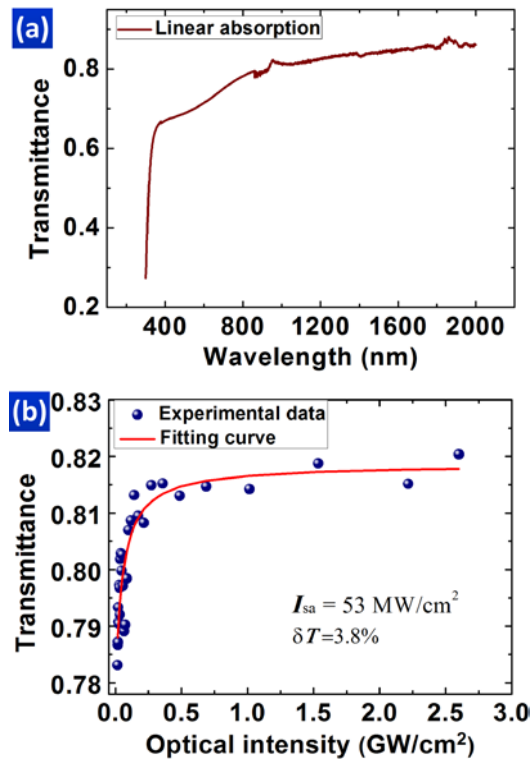

Fig.2. (a) The transmission spectrum of the few-layer TI: Bi2Se3 sample, and (b) its saturable absorption characteristic at 800 nm wavelength by the Z-scan technique.

## 3. Experimental setup

To fully exploit the broadband saturable absorption of the TI, we further designed a 1.06 μm passively Q-switched YDF laser with the few-layer TI: Bi2Se3 as a Q-switcher. The experimental setup of the proposed Q-switched YDF laser is schematically shown in Fig.3. The compact linear cavity consists of a 20-cm high-concentration YDF, two fiber Bragg gratings (FBG1 & FBG2) and the TI: Bi2Se3 Q-switcher. The total cavity length is ~3.2 m. The pump laser from a 190 mW/974 nm laser diode (LD) was directly launched into the YDF for providing the laser gain. The FBG1 and FBG2 were used to form the linear resonant cavity at the 1067 nm wavelength. The chirped FBG1 has a high reflectivity of >99.9% in the range of 1067.01~1068.20 nm, and as given in Fig.3(a), the FBG2 has two reflection wavelengths of 1067.66 and 1067.92 nm with the reflectivity of 98.0% and 97.8%, respectively. Therefore, ~2% intracavity light can be extracted as laser output from the FBG2. The output laser is measured by an optical spectrum analyzer (OSA, Advantest Q8384), an optical powermeter and a high-speed photodetector together with a 1 GHz oscilloscope (Tektronix TDS2024) or radio-frequency (RF) spectrum analyzer (Gwinstek GSP-930).

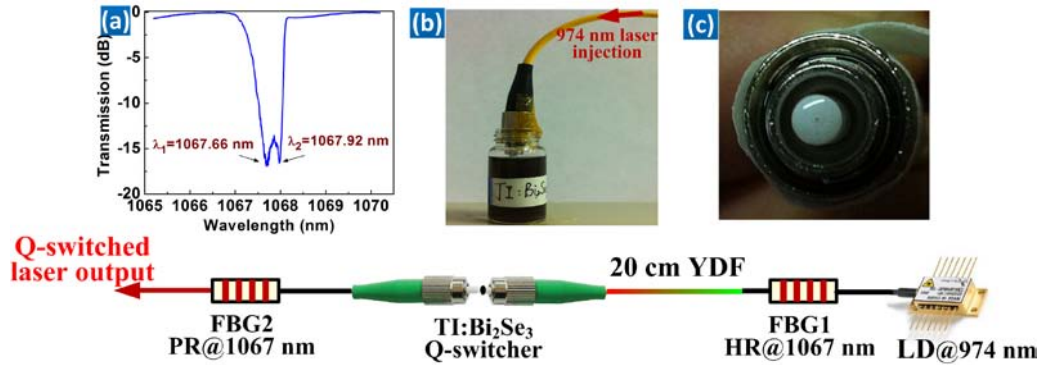

Fig.3. The experimental setup of the TI-based passively Q-switched YDF laser. (a) the transmission spectrum of the FBG2, (b) the optical deposition setup, (c) the image of the fiber-ferrule facet after depositing the TI nanosheets.

The saturable absorption of TI: Bi2Se3 plays a important role in the experiment, but the key problem is how to integrate the TI nanosheets into the fiber laser for guaranteeing the compact all-fiber structure. In this work, we used the optical deposition technique [16,17] to induce the few-layer TI nanosheets in the suspension onto a fiber ferrule. The optical-deposition setup is given in Fig.3(b). When the 20 mW laser at 974 nm was injected into the fiber, the fiber ferrule in the TI suspension started the optical deposition for about 30 minutes, and was then moved out the suspension to evaporate for 3 h at 60 °C. After the optical-deposition process, it is clearly seen from the image [Fig.3(c)] of the fiber-ferrule facet that some TI nanosheets were successfully deposited on the fiber core. At last, the TI-deposited fiber ferrule was connected with another ferrule by an optical adaptor to construct the fiber-compatible TI: Bi2Se3 Q-switcher. The insertion loss of this device is 1.9 dB at 1067 nm.

## 4. Experimental results and discussions

In the experiment, the YDF laser started the continuous-wave lasing at the pump power of 36.7 mW, and then transited to the Q-switching operation at ~42.5 mW. When the pump power was gradually increased from 42.7 to 106.2 mW, as shown in Fig.4, the stable pulse trains with the different repetition rates were observed, which is the typical feature of passive Q-switching. At the pump power of 77.7 mW, Fig.5 summarizes the typical characteristics of the Q-switching operation. The output optical spectrum in Fig.5(a) has the two lasing wavelengths of 1067.66 and 1067.91 nm, corresponding to the reflection wavelengths of the FBG2. One can see that the intensity of the weak peak (1067.91 nm) is much less (<10 dB) than that of the strong peak (1067.66 nm), and therefore the effect of the weak peak on the Q-switching can be neglected. The 3-dB bandwidth of the 1067.66-nm lasing channel is 0.10 nm. Fig.5(b) shows the oscilloscope trace of the Q-switched pulse trains. The pulse train has the period of 51.3 μs and no timing jitter is noticeable. Fig.5(c) plots the typical pulse envelope, having the symmetrical Gaussian-like shape. The pulse duration is 2.71 μs, comparable to those reported fiber lasers Q-switched with other SAs (e.g. CNTs [18], SESAM [19], graphene [16,20-22]). The RF output spectrum in Fig.5(d) shows that the Q-switched pulses have the repetition rate of 19.5 kHz which agrees with the pulse period of 51.3 μs measured in Fig.5(b). Moreover, the RF signal-to-noise ratio is as high as 48 dB (~$10^5$ contrast), and the broadband RF spectrum in the inset of Fig.5(d) shows no spectral modulation, indicating that the passively Q-switching operation was very stable.

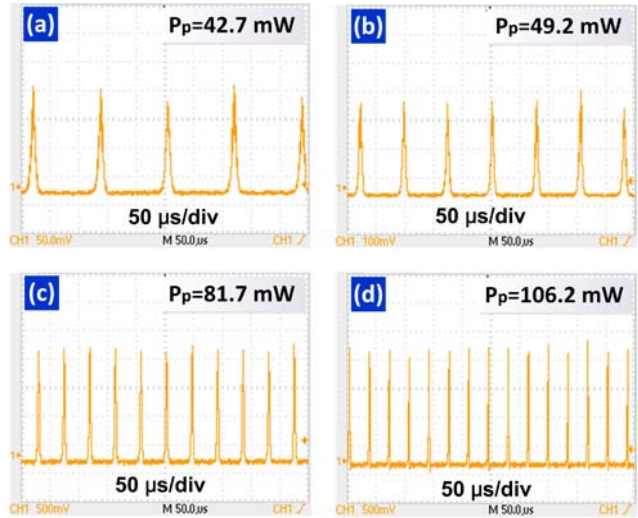

Fig.4. The Q-switched pulse trains under different pump powers $P$p, (a) $P$p =42.7 mW, (b) $P$p =49.2 mW, (c) $P$p =81.7 mW, (d) {P} $P$p =106.2 mW.

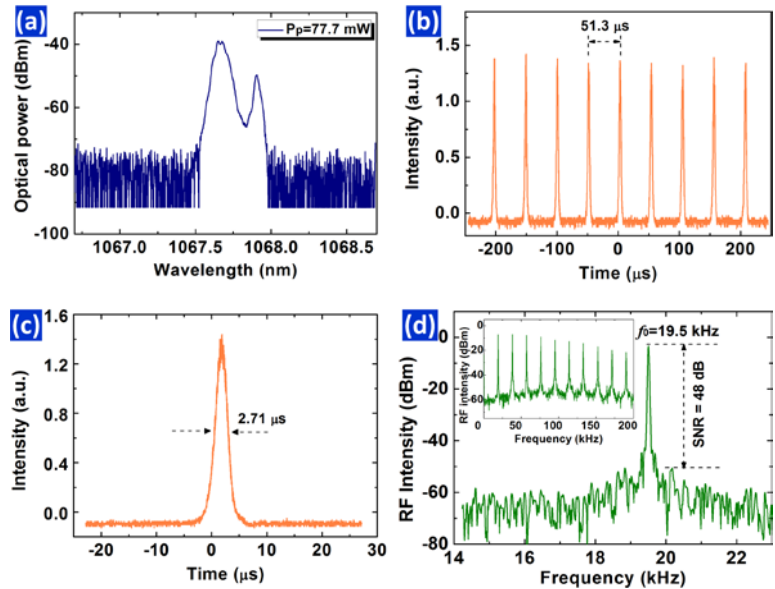

Fig.5. The typical Q-switching characteristics at the pump power of 77.7 mW. (a) output optical spectrum, (b) the oscilloscope trace of the Q-switched pulse train, (c) single pulse envelope, (d) RF output spectrum, and *Inset*: broadband RF spectrum.

As shown in Fig.6(a), we recorded the pulse repetition rate and the pulse duration as a function of the pump power. As increasing the pump power from 42.5 to 106.2 mW, the repetition rate of the Q-switching pulses can be tuned from 8.3 to 29.1 kHz, and meanwhile the pulse duration was significantly narrowed from 8.3 to 1.95 $\mu$s. The pulse duration could be further narrowed by optimizing the parameters, including

1) shortening the cavity length and 2) improving the modulation depth of the TI: Bi2Se3 Q-switcher [23]. In addition, as plotted in Fig.6(b), we also measured the average output power and correspondingly calculated the single-pulse energy. The average output power almost linearly increased with the input pump power, and at the pump power of 106.2 mW, the maximum average output power is 0.46 mW. One can see from Fig.6(b) that the pulse energy linearly grew in the initial stage, but after the pump power over 90 mW, the pulse energy became to saturate obviously. The maximum pulse energy is 17.9 nJ. The low average power and the saturated pulse energy could be attributed to the facts that the FBG2 with high reflectivity (98.0%) as the output mirror extracts the very little laser signal and the low-loss cavity limits the Q-switched pulse energy as analyzed in Ref.[23].

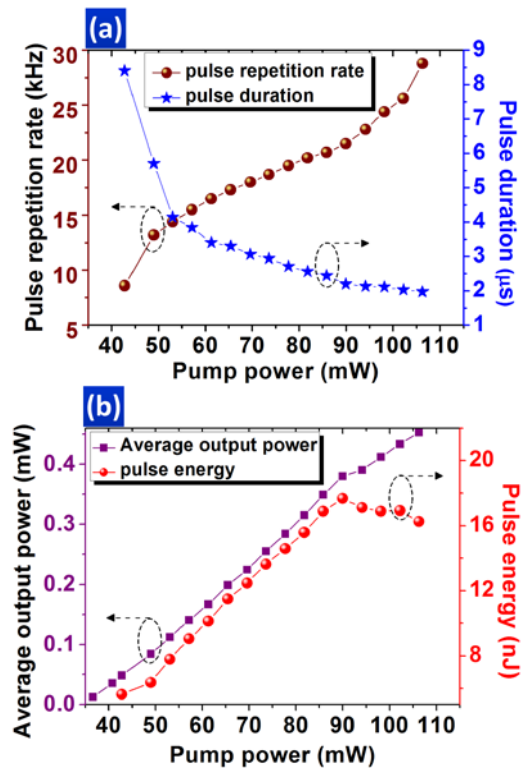

Fig.6. The pulse duration and the pulse repetition rate as a function of pump power, (b) the average output power and the pulse energy as a function of pump power.

## 5. Conclusion

In conclusion, we have proposed and experimentally demonstrated the TI-based passively Q-switched all-fiber laser operating at 1 μm waveband. Using the optical deposition method, the few-layer TI: Bi2Se3 in the suspension was well deposited onto a fiber ferrule to fabricate the fiber-compatible TI-based Q-switcher. The stable Q-switching operation at 1.067 μm was successfully achieved with the maximum pulse energy of 17.9 nJ, the shortest pulse duration of 1.95 μs and the pulse repetition rate from 8.3 to 29.1 kHz. Our results show that: 1) the Q-switching performances based on the few-layer TI are comparable to those obtained with other SAs (e.g. CNTs [18], graphene [16,20-22]) and 2) the TI nanomaterials could operate as the ultra-broadband SA.

*Acknowledgement:* This work is supported partially by the National Natural Science Foundation of China (NSFC) (No. 61177044 and 61107038), and partially supported by the Natural Science Foundation of Fujian Province of China (No. 2011J01370).